\documentclass[preprints,article,accept,pngtex,moreauthors]{Definitions/mdpi} 
\firstpage{1} 
\pubvolume{1}
\issuenum{1}
\articlenumber{0}
\pubyear{2025}
\copyrightyear{2025}
\datereceived{ } 
\daterevised{ } 
\dateaccepted{ } 
\datepublished{ } 
\hreflink{https://doi.org/} 

\Title{XANES absorption spectra of penta-graphene and penta-SiC$_2$ with different terminations: a computational study}

\TitleCitation{XANES absorption spectra of penta-graphene \\ and penta-SiC$_2$ with different terminations: a computational study}

\Author{Andrea Pedrielli $^{1}$\orcidA{}, Tommaso Morresi$^{2,3}$\orcidB{} and Simone Taioli$^{2,3}$\orcidC{}}

\AuthorNames{Andrea Pedrielli, Tommaso Morresi, Simone Taioli}

\isAPAStyle{%
       \AuthorCitation{Lastname, F., Lastname, F., \& Lastname, F.}
         }{%
        \isChicagoStyle{%
        \AuthorCitation{Lastname, Firstname, Firstname Lastname, and Firstname Lastname.}
        }{
        \AuthorCitation{Pedrielli A., Morresi T., and Taioli S.}
        }
}

\address{%
$^{1}$ Materials and Topologies for Sensor \& Devices (MTSD), Sensors and Devices Center, Fondazione Bruno Kessler, Via Sommarive 18, Povo, 38123 Trento, Italy \\ $^{2}$ European Centre for Theoretical Studies in Nuclear Physics and Related Areas (ECT*), Fondazione Bruno Kessler (FBK), Strada delle Tabarelle 286, 38122, Trento, Italy\\ $^{3}$ Trento Institute for Fundamental Physics and Applications (TIFPA), National Institute for Nuclear Physics (INFN), Via Sommarive 14, 38123, Trento, Italy}
\corres{Correspondence: pedrielli@fbk.eu}

\abstract{In recent research, penta-graphene and penta-SiC$_2$ have emerged as innovative 2D materials consisting exclusively of pentagons. However, there is still a significant gap in the theoretical characterization of these materials, which hinders progress in their synthesis and potential technological applications. This study aims to close this gap by investigating the X-ray absorption near-edge spectroscopy (XANES) of these materials through ab initio calculations. In particular, we analyze the XANES spectra of penta-graphene in its pristine, hydrogenated, and hydroxylated states, and we investigate the effects of substitution by a single silicon in both penta-graphene and pentagraphane. In addition, we calculate the XANES spectra for pristine and hydrogenated penta-SiC$_2$. This work sets the stage for the possible identification of penta-graphene and penta-SiC$_2$ phases by X-ray spectroscopy at the experimental level and lays the foundation for the future engineering of the absorption properties of these materials in optical devices.}


\begin{document}



\section{Introduction}

Among the various carbon allotropes, much attention has been devoted to graphene \citep{Geim2007}, a 2D material with remarkable mechanical \cite{pedrielli2017designing} and electronic properties \citep{Allen2010,Randviir2014,taioli2014computational,azzolini2018anisotropic,backes2020production}.
Intrinsic graphene is indeed a semimetal in which electrons and holes behave like massless fermions due to linear energy dispersion around the six Dirac corners of the Brillouin zone and exhibits high electron mobility at room temperature with reported values of over 15,000 cm$^2$V$^{-1}$s$^{-1}$ as well as high opacity for an atomic monolayer. Graphene is also a promising platform for studying the behavior of quantum systems in curved spacetime, as a connection between its low-energy Dirac-like electronic excitations and relativistic quantum field theories has been discovered when graphene is shaped into a surface with constant negative curvature, such as carbon pseudospheres with penta-heptagonal defects \cite{taioli2016lobachevsky,morresi2020exploring}. It is also one of the strongest materials known, with a breaking strength more than 100 times greater than that of a hypothetical steel foil of the same thickness \cite{morresi2020structural,pedrielli2023mechanical,pedrielli2018mechanical}. 
Since the discovery of graphene, several 2D materials \cite{carvalho2021computational} have been proposed, such as silicene \citep{Vogt2012}, boron nitride \citep{Meyer2009}, including layered heterostructures made of germanene, tinene and antimonene nanocomposites \citep{Chen2016, Matthes2013}, opening up a specific field of research in ``quantum materials''.\\ \indent Recently, a novel 2D carbon allotrope, called penta-graphene, has also been proposed \citep{Zhang2015}.
In such a material, the plane is tiled by a series of pentagons consisting of three- and four-coordinated carbon atoms. Penta-graphene is an insulator with an indirect band gap in the range of $4.1-4.3$ eV, as calculated in the G$_0$W$_0$ approximation \citep{Einollahzadeh2016,taioli2009electronic,umari2012communication}. The mechanical \citep{Le2017}, electronic and optical properties \citep{Rajbanshi2016,Yuan2017} as well as the thermal stability and thermal conductivity \citep{Xu2015,Wang2016,Wang2017} of this 2D carbon allotrope have already been characterized from a computational point of view.
In addition, the effects of hydrogen passivation on the electronic properties and magnetic moment \citep{Liu2017} as well as the mechanical properties \citep{Ebrahimi2016} were investigated. In particular, an effective tight-binding model for penta-graphene \citep{Stauber2016} was developed. The hydrogenated version of penta-graphene, the so-called pentagraphane, has an indirect band gap with a value of $5.78$ eV \citep{Einollahzadeh2016A}, which is close to the band gap of diamond. It was found that hydrogenation and fluorination can effectively tune the electronic \citep{Berdiyorov2016A} and mechanical properties of graphene \cite{haberer2010tunable,haberer2011direct} and penta-graphene \citep{Li2016}. Various strategies for the tuning of the electronic and magnetic properties of penta-graphene were proposed \citep{Santos2020Defects, Santos2020Magnetic}.
A detailed review of progress on penta-graphene and its related materials has been recently published \citep{Nazir2022}.
Although the stability of penta-graphene has been predicted theoretically, it is still controversial experimentally \citep{Ewels2015}. This does not fundamentally exclude the possibility that it can be synthesized at low temperature or stabilized by a supporting surface, as in the case of silicon nanoribbons with only pentagonal rings grown on the Ag(110) surface, which has been demonstrated both experimentally and computationally \citep{Cerda2016}. \\
\indent Similar to penta-graphene, pentagonal silicon dicarbide (penta-SiC$_2$) was recently studied using first-principles simulations \citep{Berdiyorov2016, Xu2017}.\\ 
\indent However, among the many interesting properties, a thorough analysis of the X-ray absorption spectra of these materials is still missing, which is extremely useful both for their characterization and for their technological applications \cite{taioli2010electron,taioli2024advancements}. X-ray absorption near-edge spectroscopy (XANES) has proven to be a powerful tool for gaining insight into the local chemical environment (e.g., coordination) and oxidation state of the excited atomic sites, with wide-ranging technological applications based on this excited-state property, from the characterization of catalysts \citep{Yano2009} to mediated cancer therapy \citep{Fabbri2010,Morresi2018}.\\
\indent In this work, we present a comprehensive study of the X-ray absorption of pristine, hydrogenated, and hydroxylated penta-graphene, together with an investigation of the influence of a single silicon substitution in penta-graphene and penta-graphane using accurate density functional theory (DFT) calculations. In addition, we present a similar analysis for penta-SiC$_2$ for both pristine and hydrogenated material, as summarized in Tab. \ref{table:1}. 
This study paves the way for potentially identifying penta-graphene and penta-SiC$_2$ phases by X-ray spectroscopy and also provides the basis for tailoring the absorption properties of these materials in optical devices.

\begin{table}[h!]
\centering
\caption{Summary of the examined cases for the X-ray absorption analysis}
\begin{tabularx}{0.8\textwidth}{lccc}
\toprule
Material      & Pristine &Hydrogenated & Hydroxylated \\
\hline
Penta-graphene           & yes               & yes     & yes          \\
\hline
Penta-graphene + Si sub.         & yes               & yes      & yes    \\
\hline
Penta-SiC$_2$           & yes               & yes      & ---             \\
\hline
\label{table:1}
\end{tabularx}
\end{table}

\section{Computational methods}
\subsection{Penta-graphene and penta-SiC$_2$ models}

\begin{figure}[hptb!]
\centering
\includegraphics[width=1\textwidth]{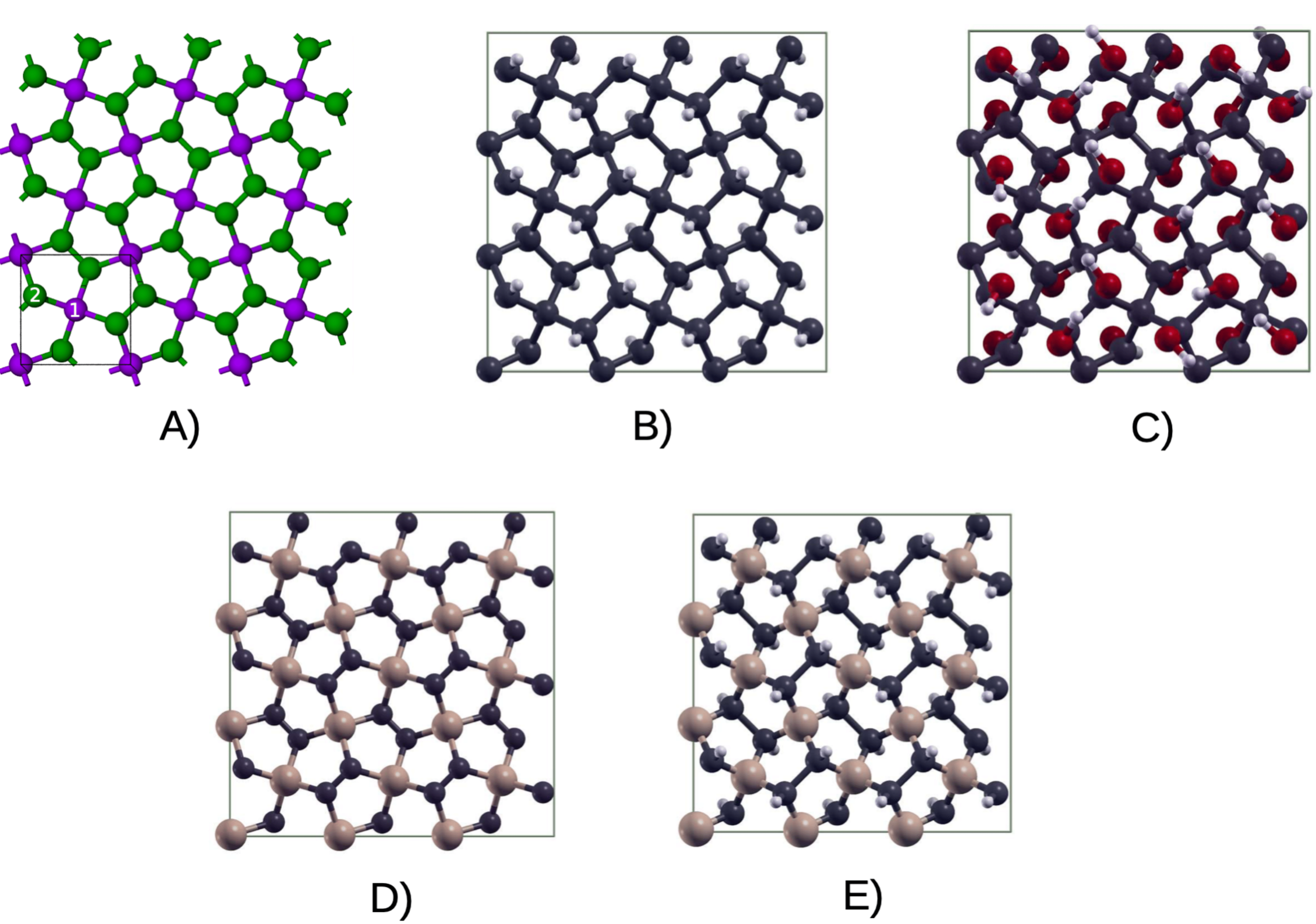}
\caption{A) 3$\times$3 supercell of penta-graphene in which tri-coordinated sites are colored green and the four-coordinated atoms are colored purple. The unit cell is framed by black lines, and the two non-equivalent sites are numbered (1) and (2). In penta-graphene, all atoms are carbon atoms, while in penta-SiC$_2$, the tri-coordinated sites are occupied by carbon atoms and the four-coordinated sites by silicon atoms. The image was produced using Ovito software \citep{Stukowski2009}. 3$\times$3 supercells of B) hydrogenated penta-graphene, C) hydroxylated penta-graphene, D) penta-SiC$_2$ and E) hydrogenated penta-SiC$_2$. The carbon atoms are shown in gray, the silicon atoms in beige, the oxygen atoms in red, and the hydrogen atoms in white. The images were created with the XCrySDen software \citep{Kokalj2003}.}
\label{fig_Structures}
\end{figure}

In Fig. 
\ref{fig_Structures}A)
we show the crystal structure of penta-graphene, which consists of alternating $sp^3$- and $sp^2$-hybridized carbon atoms, colored purple and green in the ratio $1:2$, respectively. Penta-graphene exhibits P-421m symmetry with six atoms in the tetragonal unit cell shown in Fig. \ref{fig_Structures}A) within the black lines. The penta-graphene structure is not completely flat like that of graphene, but has a protrusion of $sp^3$-bonded atoms above and below the layer plane. This is due to the non-planarity caused by the presence of $sp^3$-bonded atoms.

The $sp^3$ and $sp^2$ sites are naturally inequivalent within the unit cell, so these two sites should be naturally used as target atoms for the X-ray absorption calculations. The structure of penta-SiC$_2$ is essentially the same as that of penta-graphene, with silicon atoms in the four-coordinated positions.\\
\indent In Fig. \ref{fig_Structures}, we also show a representation of B) hydrogenated penta-graphene and C) hydroxylated penta-graphene $3\times3$ supercells. In addition, penta-SiC and hydrogenated penta-SiC are shown in panels D) and E) of the same figure, respectively. In the case of hydrogenation and hydroxylation, the $sp^2$ atoms of penta-graphene or penta-SiC$_2$ are saturated by the H and OH end groups, respectively. We would like to point out that the labels of the inequivalent sites 1 and 2 remain valid for all structures investigated, also with regard to the absorption spectra below.

\indent As we will explain in more detail below, we used a 3$\times$3 supercell in all calculations of the absorption spectra, which proved to be sufficient for the convergence of the spectra with respect to the supercell dimension. 
Along the direction perpendicular to the penta-graphene (or penta-SiC$_2$) plane, we chose a supercell dimension of $2.0$ nm to avoid spurious interactions between the periodic replicas. To optimize the structure, we relaxed both the dimensions of the supercells and the positions of the atoms using DFT calculations.
The geometric minimization has been carried out with the {\sc Quantum ESPRESSO} suite \citep{Giannozzi2009}. Self-consistency was achieved using the total energy and Hellman–Feynman force criteria below $0.00136$ eV and $0.0136$ eV \AA$^{-1}$, respectively. At the same time, the dimensions of the supercell were relaxed along the planar directions until a pressure of $0.5$ kbar was reached. 

\subsection{Absorption spectra calculations}

Ab initio calculations of the XANES spectra were performed with the XSpectra program \citep{Gougoussis2009}, which is included in the {\sc Quantum ESPRESSO} suite \citep{Giannozzi2009}. The Lanczos chain algorithm implemented in the code allows the calculation of the K-edge absorption spectra based on the ground state electron density, avoiding the explicit calculation of the empty states.
The core-hole created by the absorption of X-rays is included in the pseudopotential. In particular, for the treatment of ionic core-valence electron interactions in C and Si atoms, we have used the {\sc Quantum ESPRESSO} ultra-soft core-hole pseudopotentials database \citep{Quantum}. The Perdew--Burke--Ernzerhof (PBE) \cite{pbe} was used for the exchange-correlation potential. 

The projector augmented wave (PAW) formalism \citep{Blochl1994} allows the recovery of the all-electron wave function. In particular, for the C pseudopotential we have used two projectors for the $s$ and $p$ atomic states, while for Si and O we have used two projectors for the $s$ and $p$ states and one for the $d$ orbital. After convergence tests, DFT simulations were carried out with a cut-off value for the kinetic energy of 544 ~eV and four times as much for the electron density.
A 1 $\times$ 1 $\times$ 1 $k$-point sampling of the Brillouin zone was sufficient to obtain converged energy and DOS in the case of a 3$\times$3 simulation cell, while a 2$\times$2 simulation cell required a 2 $\times$ 2 $\times$ 1 $k$-point sampling. 
XANES spectra are calculated in the dipole approximation; therefore, only the transitions from $1s$ to $np$ or $s/p$ hybridized orbitals are examined in K-edge spectroscopy due to symmetry constraints imposed by the selection rules.
The spectrum simulations were performed using a 4 $\times$ 4 $\times$ 1 $k$-point grid to sample the Brillouin zone. Three different polarization vectors of the incident light were considered, namely one perpendicular to the penta-graphene and penta-SiC$_2$ plane ($z$-axis) and two parallel to it and perpendicular to each other ($x$- and $y$-direction). In all XANES simulations, the spectral lines were broadened by a convolution with a Lorentzian function whose full width at half maximum is $0.8$~eV at all energies.

\section{Results and Discussion}

\subsection{Penta-graphene}

\begin{figure}[hptb!]
\centering
\includegraphics[width=0.5\textwidth]{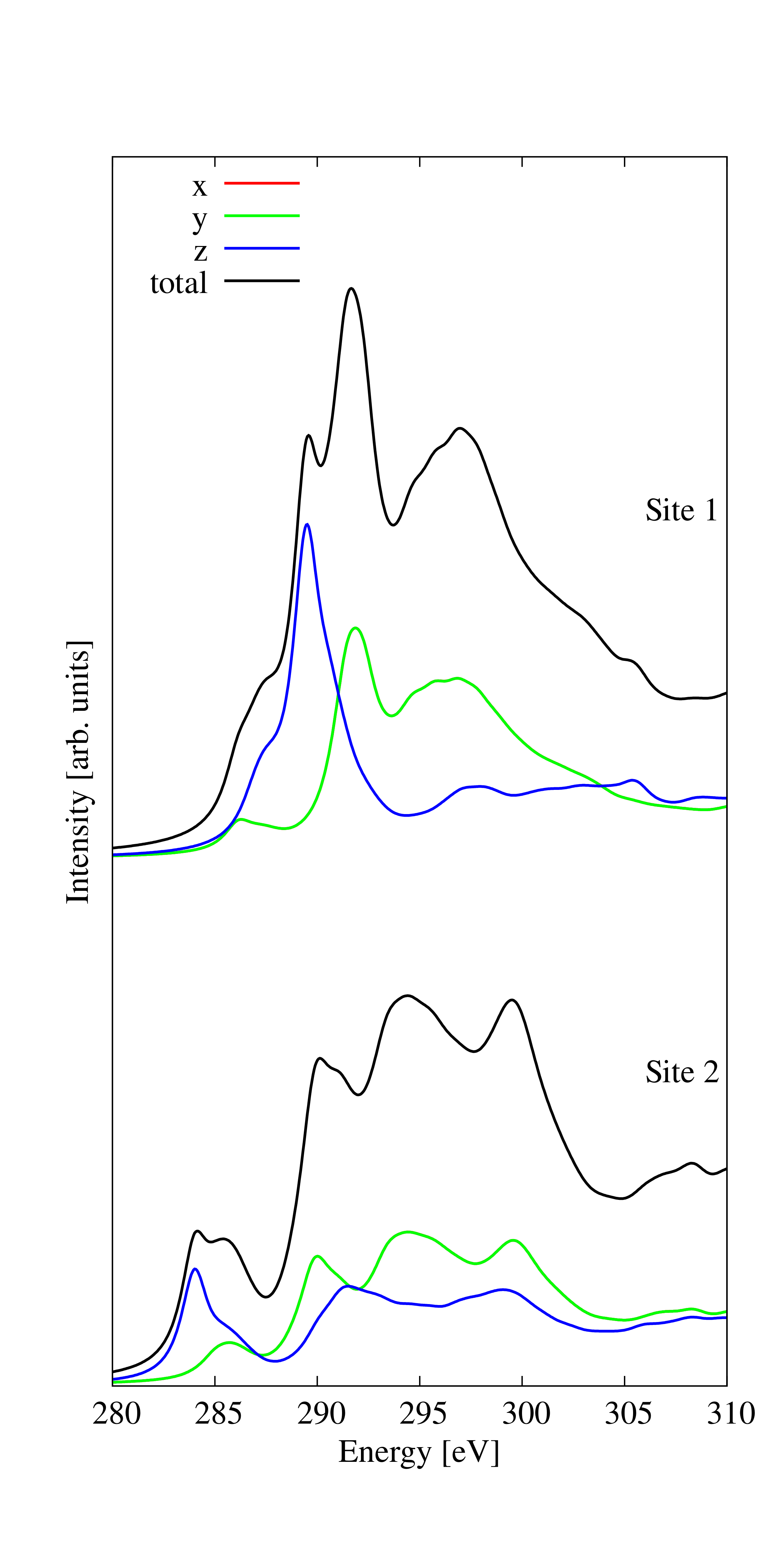}
\caption{Theoretical C K-edge XANES spectra of a 3$\times$3 unit cell of penta-graphene for inequivalent positions of the core-hole within the unit cell, for $x$-, $y$- and $z$-polarized electric fields. We note that the spectrum for $x$-polarized X-rays (red line, not visible in the diagram) is the same as that for $y$-polarized light (green line).}
  \label{fig_PentaCarbonTriple}
\end{figure}

\begin{figure}[hptb!]
\centering
\includegraphics[width=0.5\textwidth]{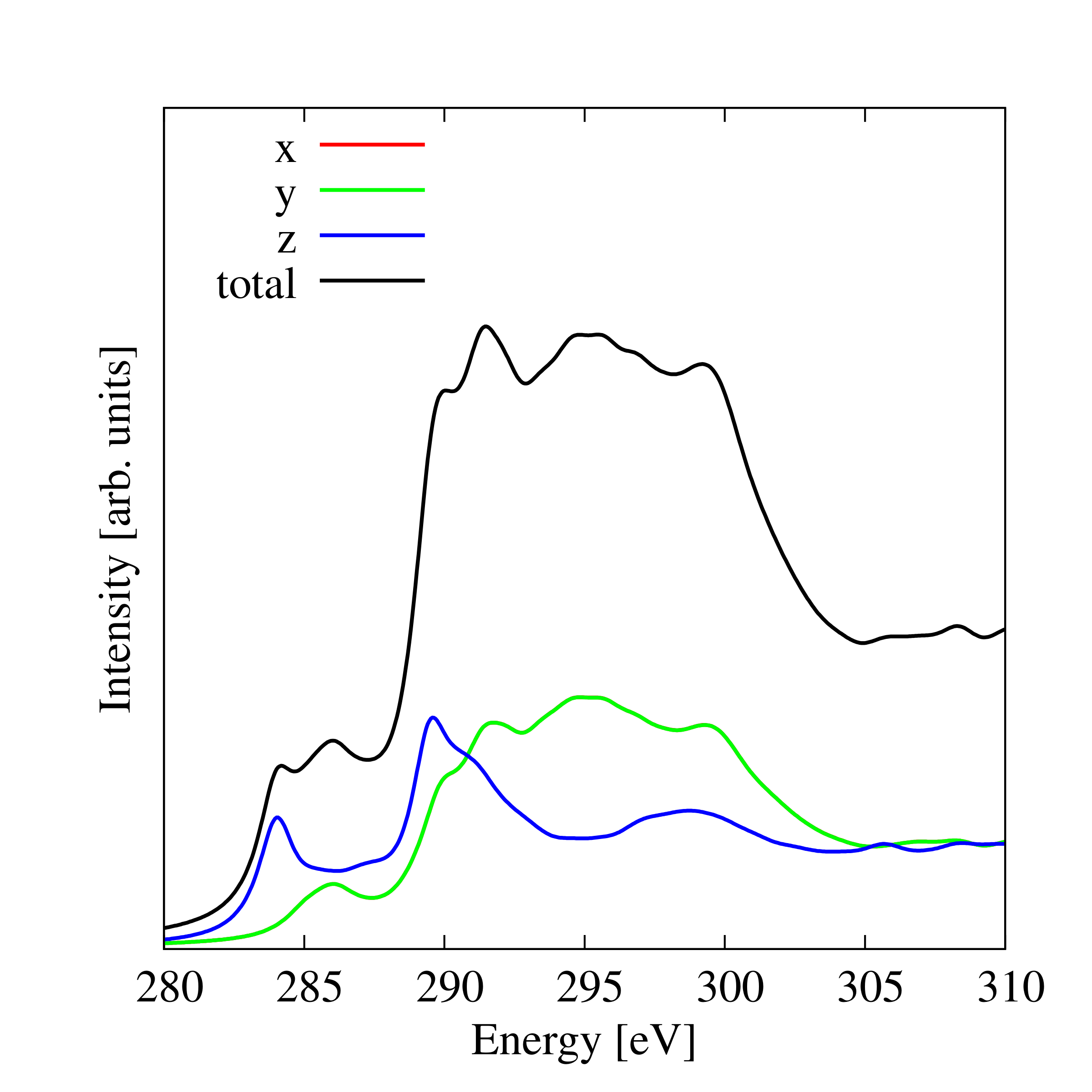}
\caption{Theoretical C K-edge XANES spectra of a 3$\times$3 unit cell of penta-graphene, averaged over all inequivalent positions of the core-hole within the unit cell, for $x$-, $y$- and $z$-polarized electric fields. We note that the spectrum for $x$-polarized X-rays (red line, not visible in the diagram) is the same as that for $y$-polarized light (green line).}
\label{fig_PentaAveragedCarbonTriple}
\end{figure}

The calculated C K-edge XANES spectra for pristine penta-graphene (3$\times$3 supercell) are shown in Fig. \ref{fig_PentaCarbonTriple}. In particular, we have calculated the spectra for the two inequivalent carbon atoms. The spectra in the upper panel of Fig. \ref{fig_PentaCarbonTriple} refer to four-coordinated carbon atoms (site 1 in Fig. \ref{fig_Structures}A), while the spectra in the lower panel were determined for a three-coordinated carbon atom (site 2 in Fig. \ref{fig_Structures}A). The peak around $285$ eV, which only occurs in site 2, can be explained by $\pi$ bonds resulting from the overlap of the $p$ orbitals of the bonded atoms. It is interesting to see how the subsequent peaks in the spectra for site 2 are the same as for graphene. Indeed, three peaks at $293$ eV, $298$ eV and $303$ eV are present in graphene, indicating transitions into the three $\sigma$ bands \citep{Pacil2008, Papagno2009, Xu2015, Christiansen2023}. We note that wherever the spectrum for $x$-polarized X-rays is not visible, the line coincides with that for $y$-polarized X-rays due to local symmetry.
We have also calculated the weighted average of the spectra for all inequivalent positions of the target atom within the unit cell; this is the total spectrum, which is to be compared with the experimental measurements and shown in Fig. \ref{fig_PentaAveragedCarbonTriple} with black lines. 
We note that the main peak at low energy is obtained for both sites in response to X-rays radiation polarized along the $z$-axis. The responses for X-rays polarized along the planar direction are instead found at higher energies.
The total spectra reported in Fig. \ref{fig_PentaAveragedCarbonTriple} can be recognized as a composition of three plateaus in the ranges $283-288$ eV, $290-300$ eV and above $305$ eV.

\subsection{Hydrogenated penta-graphene}

\begin{figure}[hptb!]
\centering
\includegraphics[width=0.5\textwidth]{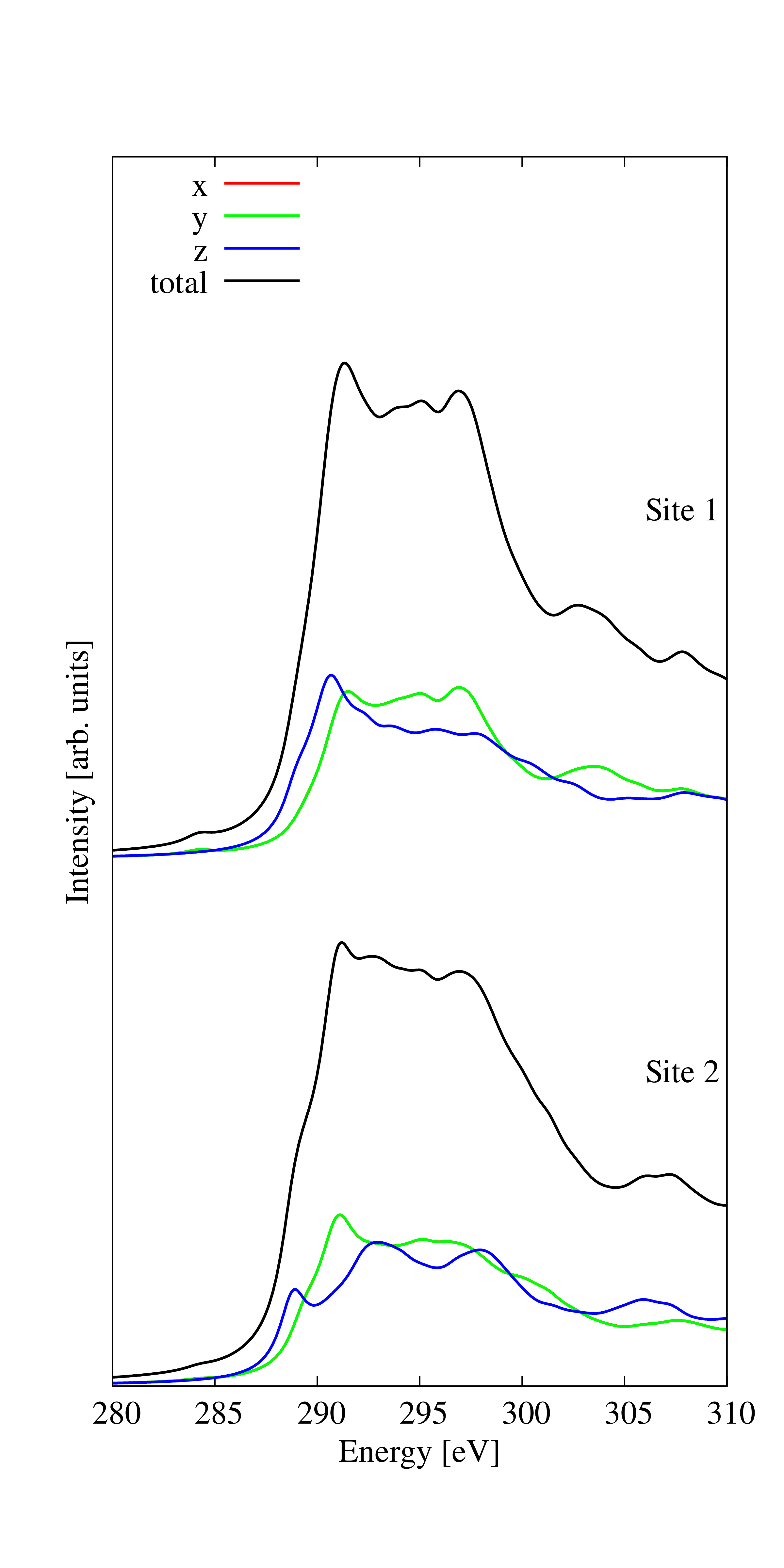}
\caption{Theoretical C K-edge XANES spectra of a 3$\times$3 supercell of hydrogenated penta-graphene for inequivalent positions of the core-hole within the unit cell, for $x$-, $y$-, and $z$-polarized electric fields. We note that the spectrum for $x$-polarized X-rays (red line, not visible in the diagram) is the same as that for $y$-polarized light (green line).}
\label{fig_HydrogenatedPentaCarbonTriple}
\end{figure}

We also analyzed the XANES response in hydrogenated penta-graphene, which has a hydrogen termination at all three-coordinated carbon atoms, leading to all four-coordinated carbon centers. We report the calculated C K-edge XANES spectra for hydrogenated penta-graphene (3$\times$3 supercell) in Fig. \ref{fig_HydrogenatedPentaCarbonTriple}. In particular, we have calculated the spectra for the two inequivalent positions of the target carbon atoms within the unit cell.
The spectra in the upper panel of Fig. \ref{fig_HydrogenatedPentaCarbonTriple} refer to a carbon atom at position 1, while the lower spectra for a hydrogen-terminated carbon atom at position 2 in Fig. \ref{fig_Structures}B (the labels refer to those in Fig. \ref{fig_Structures}A).

\begin{figure}[hptb!]
\centering
\includegraphics[width=0.5\textwidth]{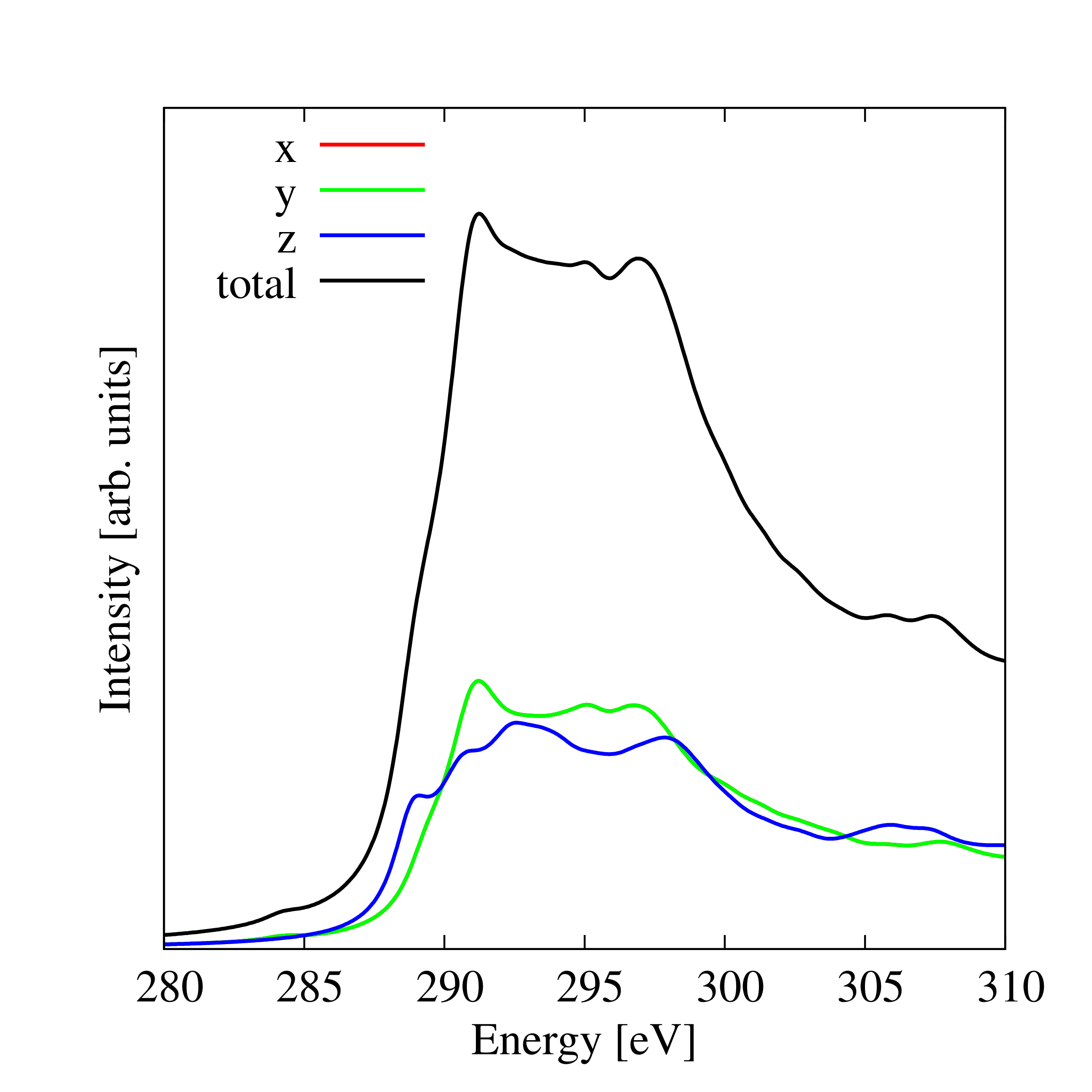}
\caption{Theoretical C K-edge XANES spectra of a 3$\times$3 supercell of hydrogenated penta-graphene averaged over all inequivalent positions of the core-hole within the unit cell, for $x$-, $y$- and $z$-polarized electric fields. We note that the spectrum for $x$-polarized X-rays (red line, not visible in the diagram) is the same as that for $y$-polarized light (green line).}
\label{fig_PentaAveragedHydrogenCarbonTriple}
\end{figure}

The hydrogen termination of the three-coordinated atoms of penta-graphene has a strong influence on the low-energy part of the XANES spectra. In addition, hydrogenation affects the relative contribution of different X-ray polarizations to the absorption spectra.
In particular, we note that i) the absorption plateau in the $283-288$ eV range that we found in the pristine case (Fig. \ref{fig_PentaAveragedCarbonTriple}) is not as pronounced in Fig. \ref{fig_PentaAveragedHydrogenCarbonTriple} and ii) the X-ray absorption spectra depend only weakly on the polarization.

\subsection{Hydroxylated penta-graphene}

\begin{figure}[hptb!]
\centering
\includegraphics[width=0.5\textwidth]{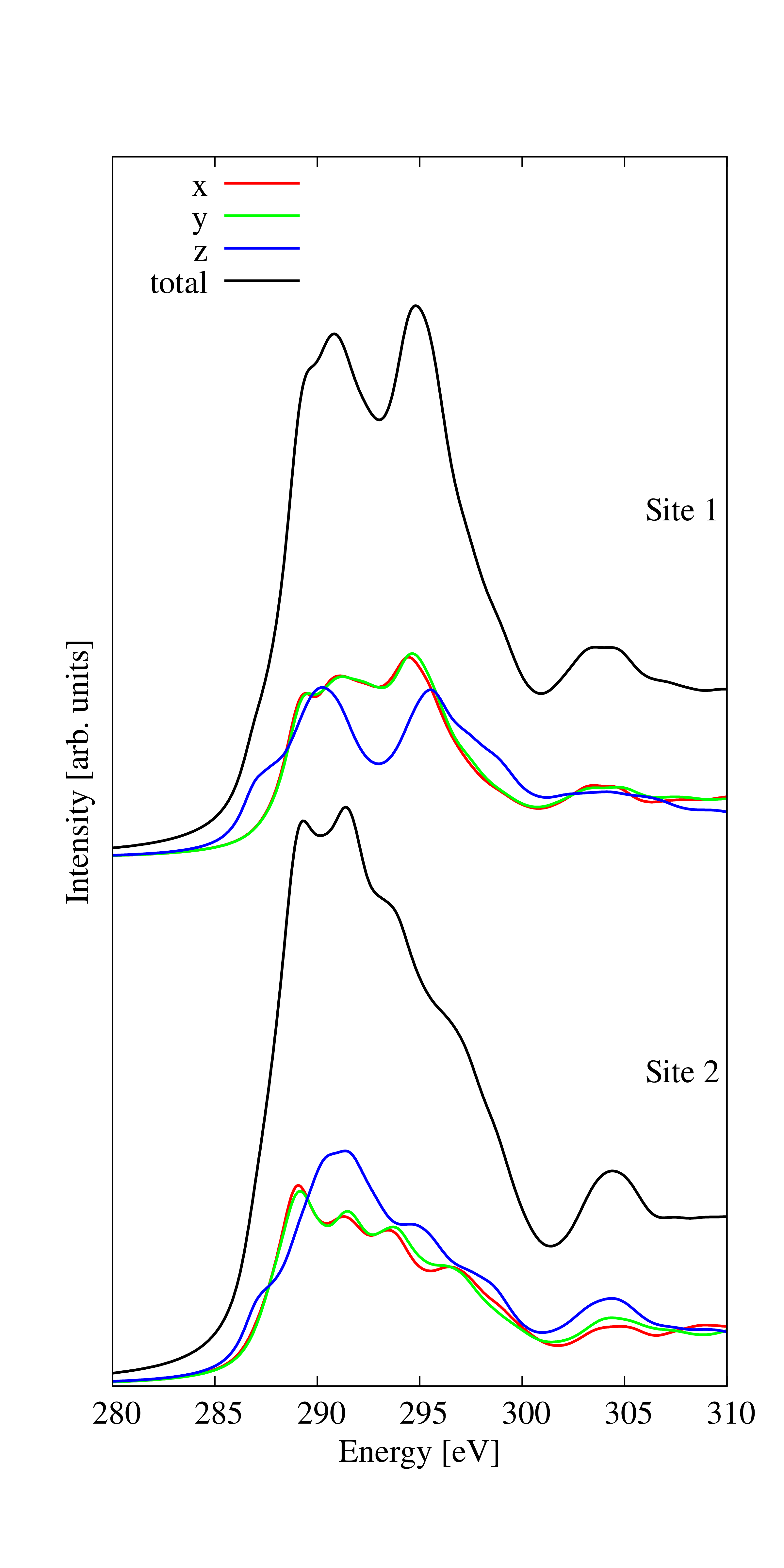}
\caption{Theoretical C K-edge XANES spectra of a 3$\times$3 supercell of hydroxylated penta-graphene for inequivalent positions of the core-hole within the unit cell, for $x$-, $y$- and $z$-polarized electric fields.}
\label{fig_HydroxylatedPentaCarbonTriple}
\end{figure}

\begin{figure}[hptb!]
\centering
\includegraphics[width=0.5\textwidth]{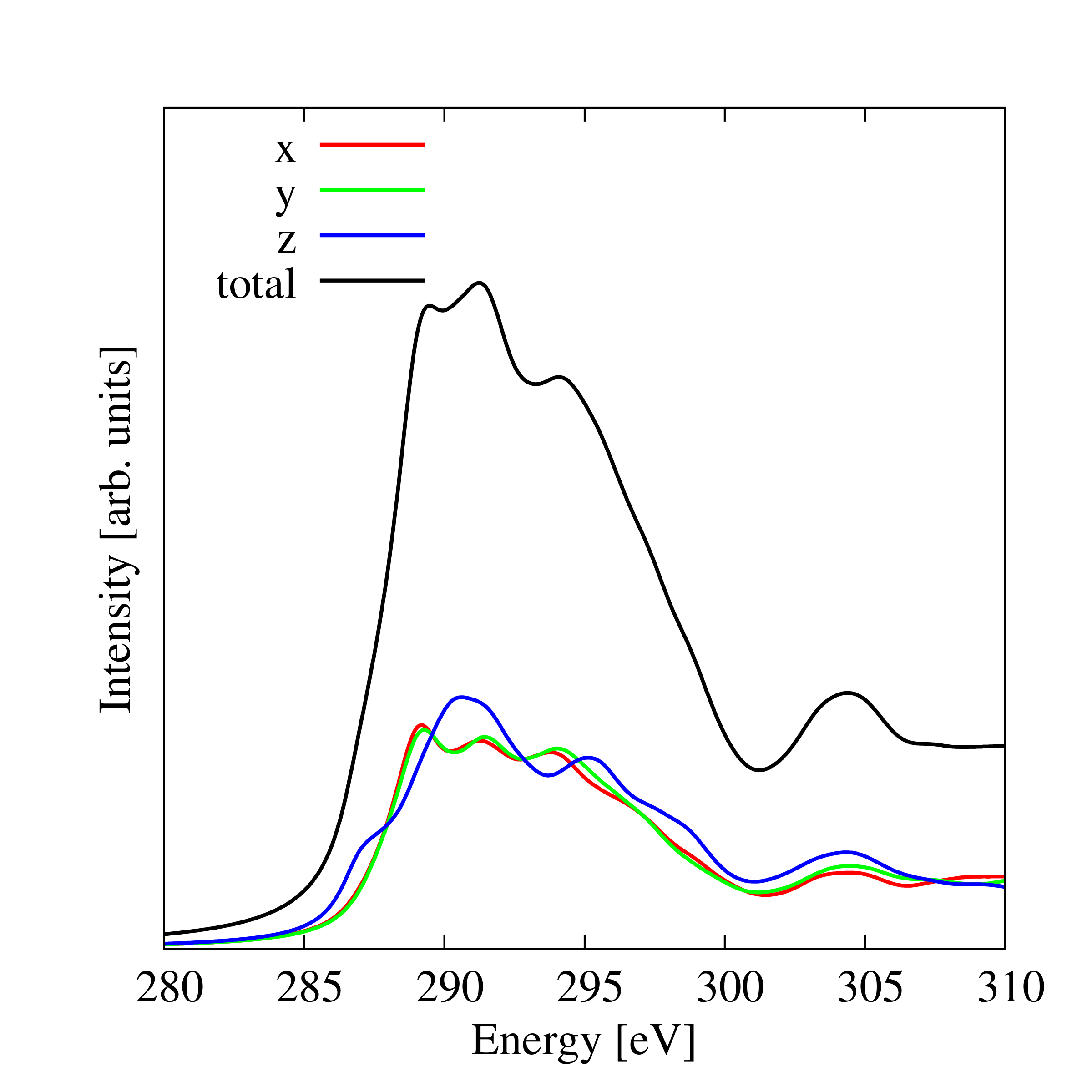}
\caption{Theoretical C K-edge XANES spectra of a 3$\times$3 supercell of hydroxylated penta-graphene averaged over all inequivalent positions of the core-hole within the unit cell, for $x$-, $y$- and $z$-polarized electric fields.}
\label{fig_PentaAveragedHydroxylatedCarbonTriple}
\end{figure}

\begin{figure}[hptb!]
\centering
\includegraphics[width=0.5\textwidth]{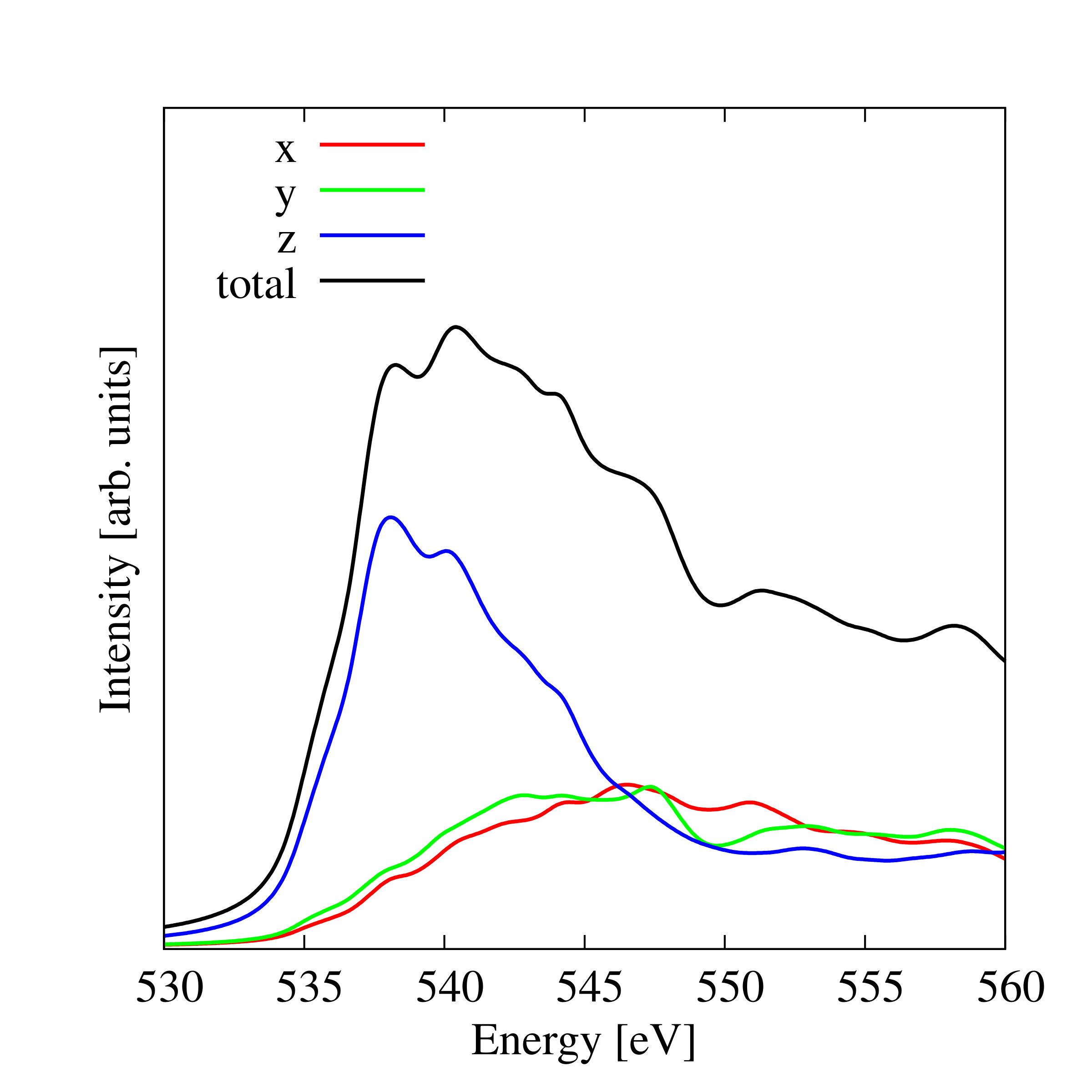}
\caption{Theoretical O K-edge XANES spectra of a 3$\times$3 supercell of hydroxylated penta-graphene for $x$-, $y$- and $z$-polarized electric fields.}
\label{fig_HydroxylatedPentaOxygenTriple}
\end{figure}

We analyzed the XANES response in hydroxylated penta-graphene. In this case, all atoms except the hydrogen and oxygen atoms are four-coordinated. In Fig. \ref{fig_HydroxylatedPentaCarbonTriple}, the calculated C K-edge XANES spectra for hydroxylated penta-graphene (3$\times$3 supercell) are shown. In particular, we have calculated the spectra for the two inequivalent positions of the target carbon atoms within the unit cell. The upper spectra in Fig. \ref{fig_HydroxylatedPentaCarbonTriple} refer to a carbon atom at site 1 as in Fig. \ref{fig_Structures}A, while the lower spectra are for a carbon atom at site 2 in Fig. \ref{fig_Structures}A. 
In addition, we calculated the weighted average of the spectra for all inequivalent positions of the target atom within the unit cell; this total spectrum is the one to be compared with the experimental measurements. We show these spectra in Fig. \ref{fig_PentaAveragedHydroxylatedCarbonTriple}.
As in the case of hydrogenated penta-graphene, the plateau at $283-288$ eV that we found in the pristine case (see Fig. \ref{fig_PentaAveragedCarbonTriple}), is not visible in Fig. \ref{fig_PentaAveragedHydroxylatedCarbonTriple}. The total spectra for inequivalent polarization directions (in-plane and out-of-plane) are very similar.
The O K-edge XANES spectra shown in Fig. \ref{fig_HydroxylatedPentaOxygenTriple} are strongly dependent on the polarization direction. In particular, most of the total spectra at low energy can be attributed to the response to $z$-polarized X-rays. 

\subsection{Silicon substitution in pristine penta-graphene}

\begin{figure}[hptb!]
\centering
\includegraphics[width=0.5\textwidth]{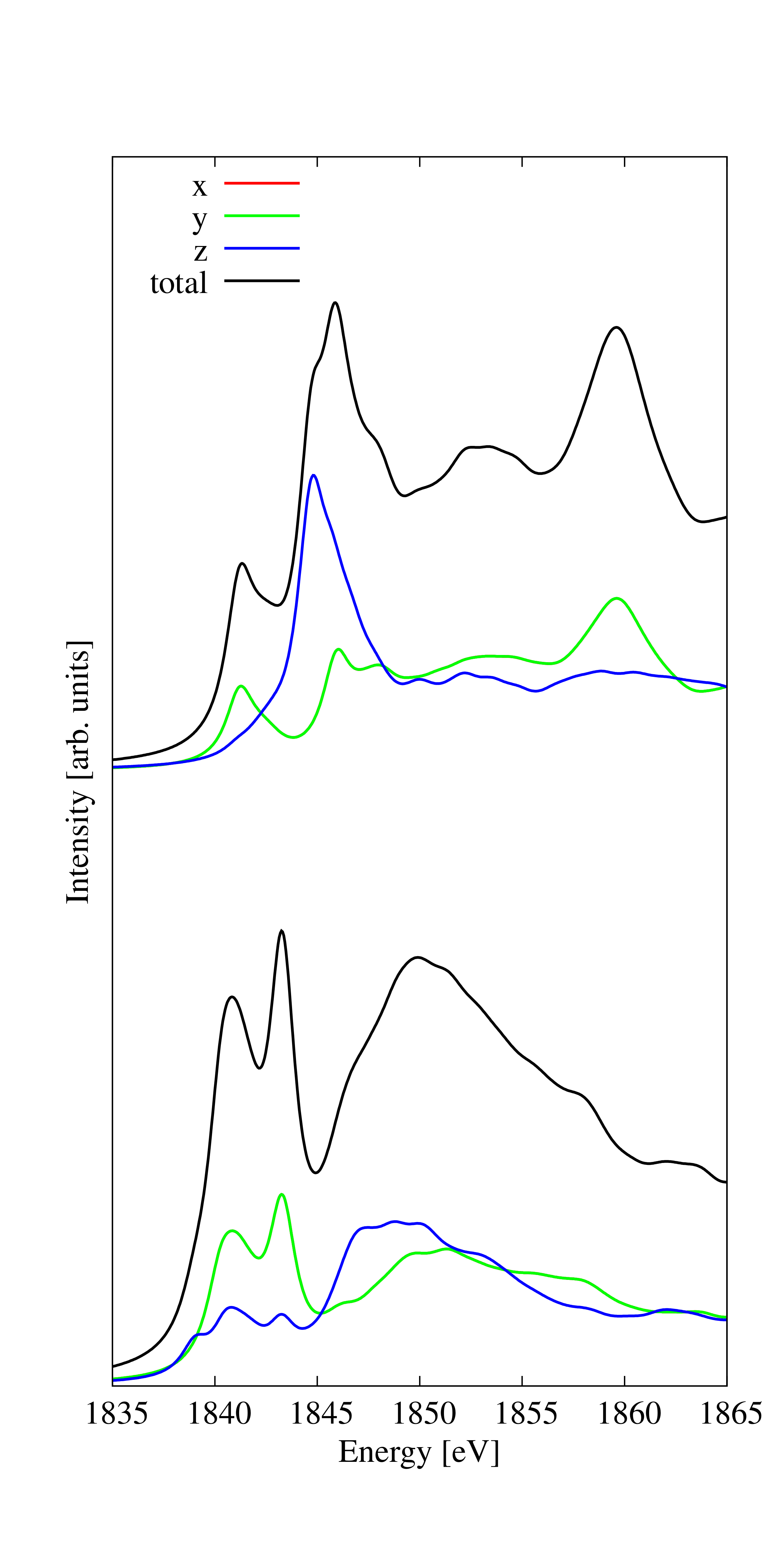}
\caption{Theoretical Si K-edge XANES spectra of a 3$\times$3 supercell of penta-graphene for inequivalent positions of the core-hole within the unit cell, for $x$-, $y$- and $z$-polarized electric fields. We note that the spectrum for $x$-polarized X-rays (red line, not visible in the diagram) is the same as that for $y$-polarized light (green line).}
\label{fig_PentaSiliconTriple}
\end{figure}

The calculated Si K-edge XANES spectra for silicon-doped penta-graphene (3$\times$3 supercell) are shown in Fig. \ref{fig_PentaSiliconTriple}. As in the case of carbon, we have calculated the spectra for the two inequivalent positions of the target silicon atom within the unit cell. If the silicon atom is in the four-coordinated site, the total spectrum (upper panel in Fig. \ref{fig_PentaSiliconTriple}) shows two main peaks at $1844$ and $1857$ eV, which are separated by a flat plateau. If the silicon atom is located at the three-coordinated site, the total spectrum (lower panel in Fig. \ref{fig_PentaSiliconTriple}) again shows two main peaks, the first in the $1840-1845$ eV range, which is made by two different contributions of the polarization of the electric field, and the second broad peak in the $1845-1860$ eV range.
We find a clear dependence on the polarization of the X-rays for both locations of Si.

\subsection{Silicon substitution in hydrogenated penta-graphene}

\begin{figure}[hptb!]
\centering
\includegraphics[width=0.5\textwidth]{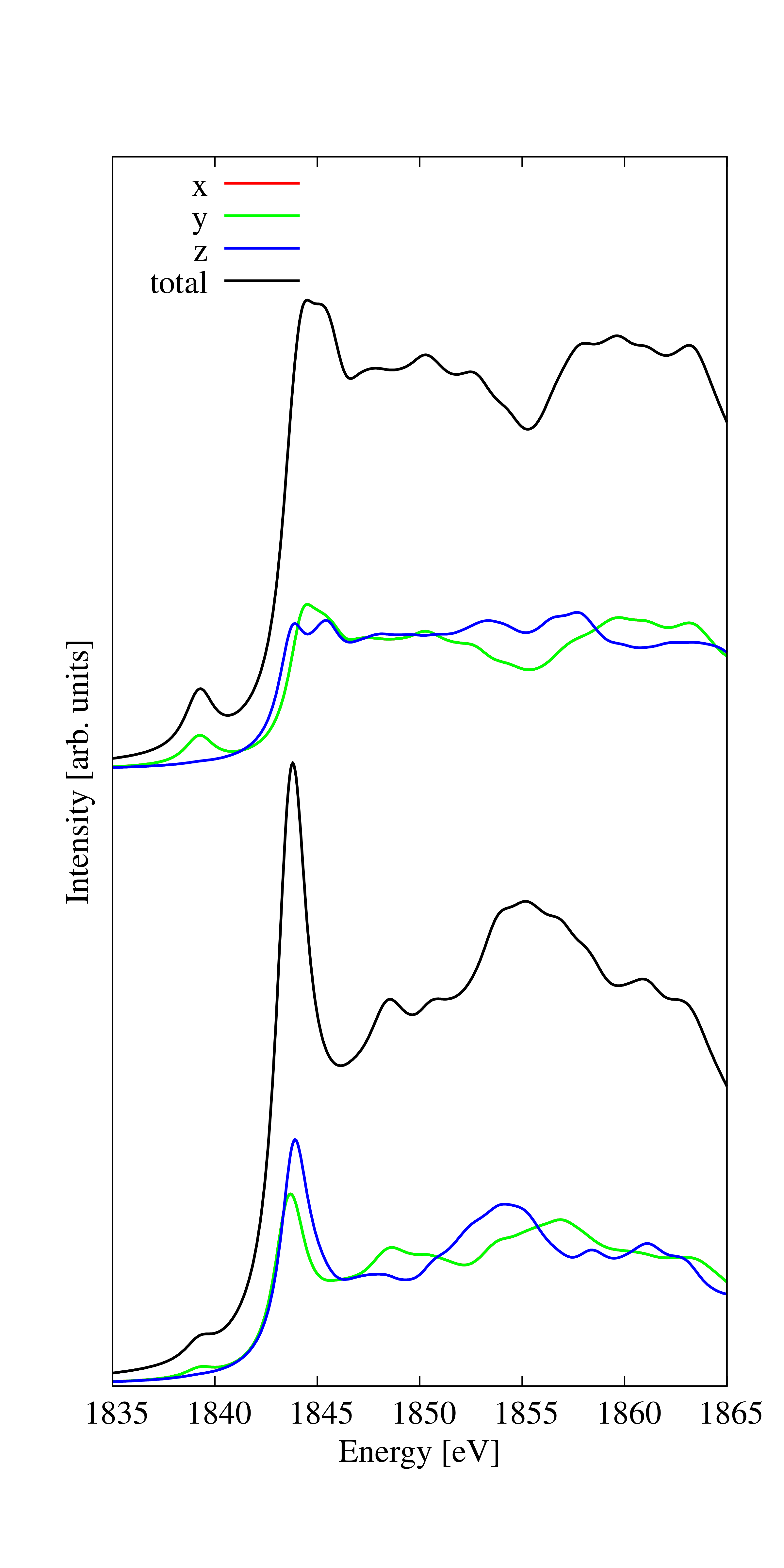}
\caption{Theoretical Si K-edge XANES spectra of a 3$\times$3 supercell of hydrogenated penta-graphene for different positions of the core hole within the unit cell, for $x$-, $y$- and $z$-polarized electric fields. We note that the spectrum for $x$-polarized X-rays (red line, not visible in the diagram) is the same as that for $y$-polarized light (green line).}
\label{fig_HydrogenatedPentaSiliconTriple}
\end{figure}

With respect to the single silicon substitution in hydrogenated penta-graphene, we report the Si K-edge XANES spectra (3$\times$3 supercell) in Fig. \ref{fig_HydrogenatedPentaSiliconTriple}. We have calculated the spectra for the two inequivalent positions of the target silicon atom within the unit cell. The spectra for the different substitution sites, as in the previous cases, site 1 and site 2 in Fig. \ref{fig_Structures}A, show a weak dependence on the polarization direction of the X-rays.

If the silicon substitution in hydrogenated penta-graphene is at site 1, the partial and the total spectrum (Fig. \ref{fig_HydrogenatedPentaSiliconTriple} upper panel) show a plateau beyond $1844$ eV. In the case where the silicon atom is located at site 2, it has three carbon atoms and one hydrogen atom as its closest neighbors. In this case, the spectra are again almost independent of the polarization and show a sharp absorption peak at $1843-1844$ eV (Fig. \ref{fig_HydrogenatedPentaSiliconTriple} lower panel).

\subsection{Penta-SiC$_2$}

Our analysis was extended to the X-ray absorption of pristine penta-SiC$_2$. The two inequivalent sites 1 and 2 in Fig. \ref{fig_Structures}A are now occupied by a silicon and a carbon atom, respectively.

\begin{figure}[hptb!]
\centering
\includegraphics[width=0.5\textwidth]{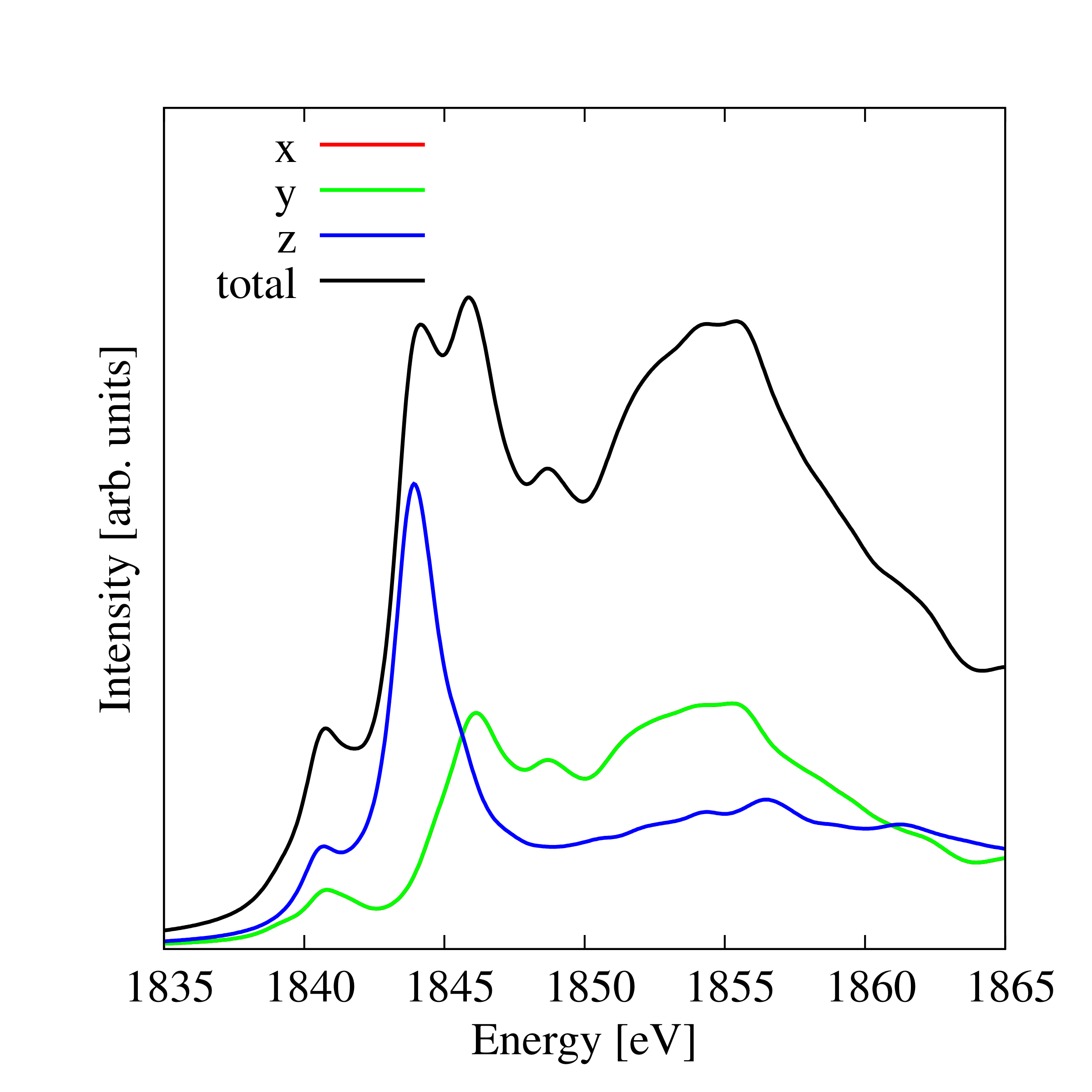}
\caption{Theoretical Si K-edge XANES spectra of a 3$\times$3 supercell of penta-SiC$_2$ with the core-hole at site 1 for $x$-, $y$- and $z$-polarized electric fields. We note that the spectrum for $x$-polarized X-rays (red line, not visible in the diagram) is the same as that for $y$-polarized light (green line).}
\label{fig_PentaSiCSiliconTriple}
\end{figure}

\begin{figure}[hptb!]
\centering
\includegraphics[width=0.5\textwidth]{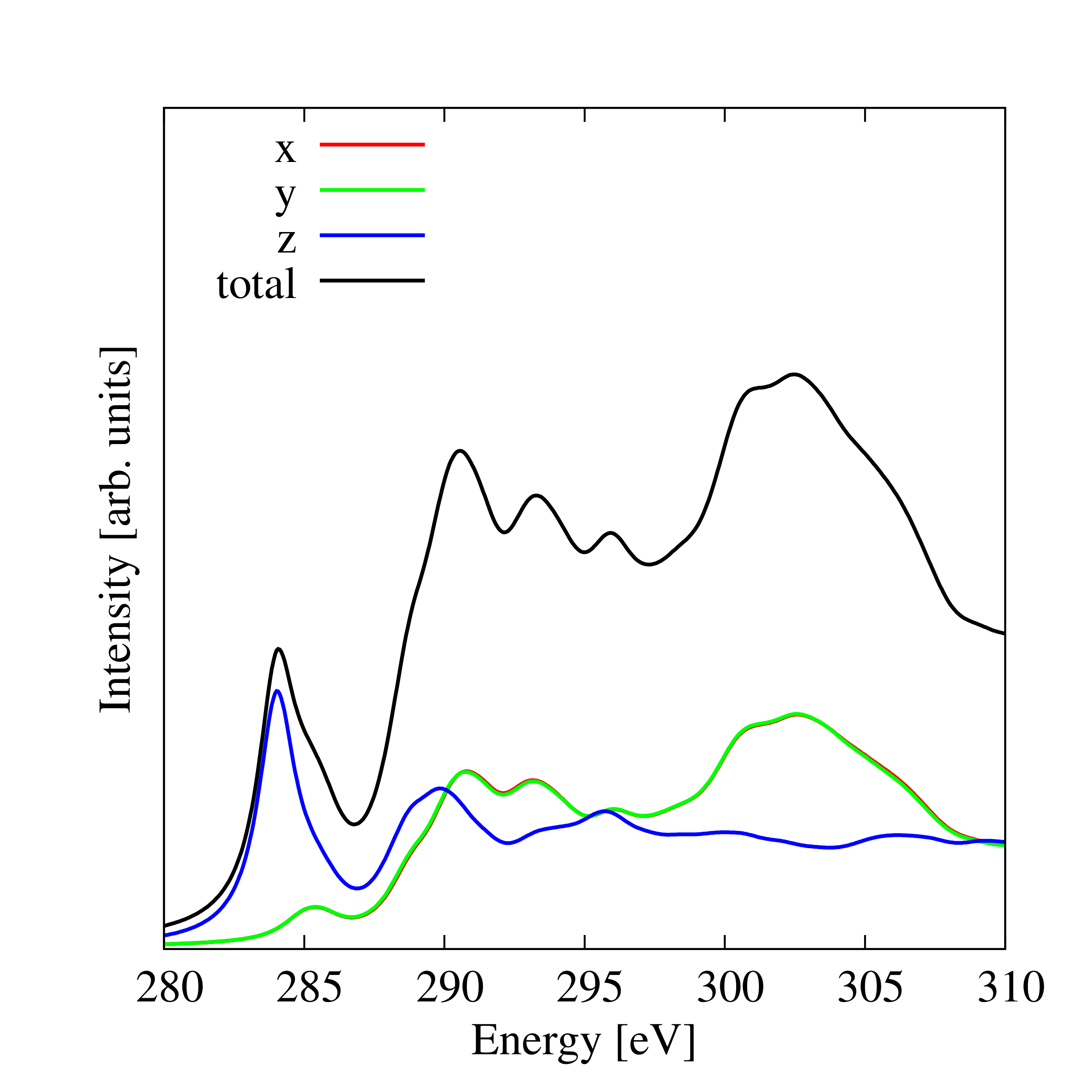}
\caption{Theoretical C K-edge XANES spectra of a 3$\times$3 supercell of penta-SiC$_2$ with the core-hole at site 2 for $x$-, $y$- and $z$-polarized electric fields. We note that the spectrum for $x$-polarized X-rays (red line, not visible in the diagram) is the same as that for $y$-polarized light (green line).}
\label{fig_PentaSiCCarbonTriple}
\end{figure}

We report in Fig. \ref{fig_PentaSiCSiliconTriple} the Si K-edge XANES spectra of a 3$\times$3 supercell of penta-SiC$_2$ and in Fig. \ref{fig_PentaSiCCarbonTriple} the spectra for the C K-edge. 
We show in Fig. \ref{fig_PentaSiCSiliconTriple} the Si K-edge XANES spectra of a 3$\times$3 supercell of penta-SiC$_2$ and in Fig. \ref{fig_PentaSiCCarbonTriple} the spectra for the C K-edge.
It is interesting to compare the penta-SiC$_2$ Si K-edge spectra reported in Fig. \ref{fig_PentaSiCSiliconTriple} with the case of a single four-coordinated silicon substitution in penta-graphene, for which the Si K-edge XANES spectra were presented in Fig. \ref{fig_PentaSiliconTriple}. At low energy, the main difference is the decoupling of the peaks in the low-energy main peak around $1845$ eV due to the shift of the $z$-polarized contribution to the spectra towards low energy. However, the most important change, which is due to the change of the atomic environment from penta-graphene to penta-SiC$_2$, is found at energies beyond $1850$ eV. The high-energy peak in the Si K-edge spectra for silicon substitution in penta-graphene is essentially suppressed, while a broad peak occurs between $1850$ and $1860$ eV.
With respect to the C K-edge results (Fig. \ref{fig_PentaSiCCarbonTriple}), the system shows a clear anisotropic behavior with the main contribution of $z$-polarized X-rays at low energy, while in the $300-305$ eV range the main contribution can be attributed to the absorption of X-rays polarized along the in-plane directions.

\subsection{Hydrogenated penta-SiC$_2$}

\begin{figure}[h!]
\centering
\includegraphics[width=0.5\textwidth]{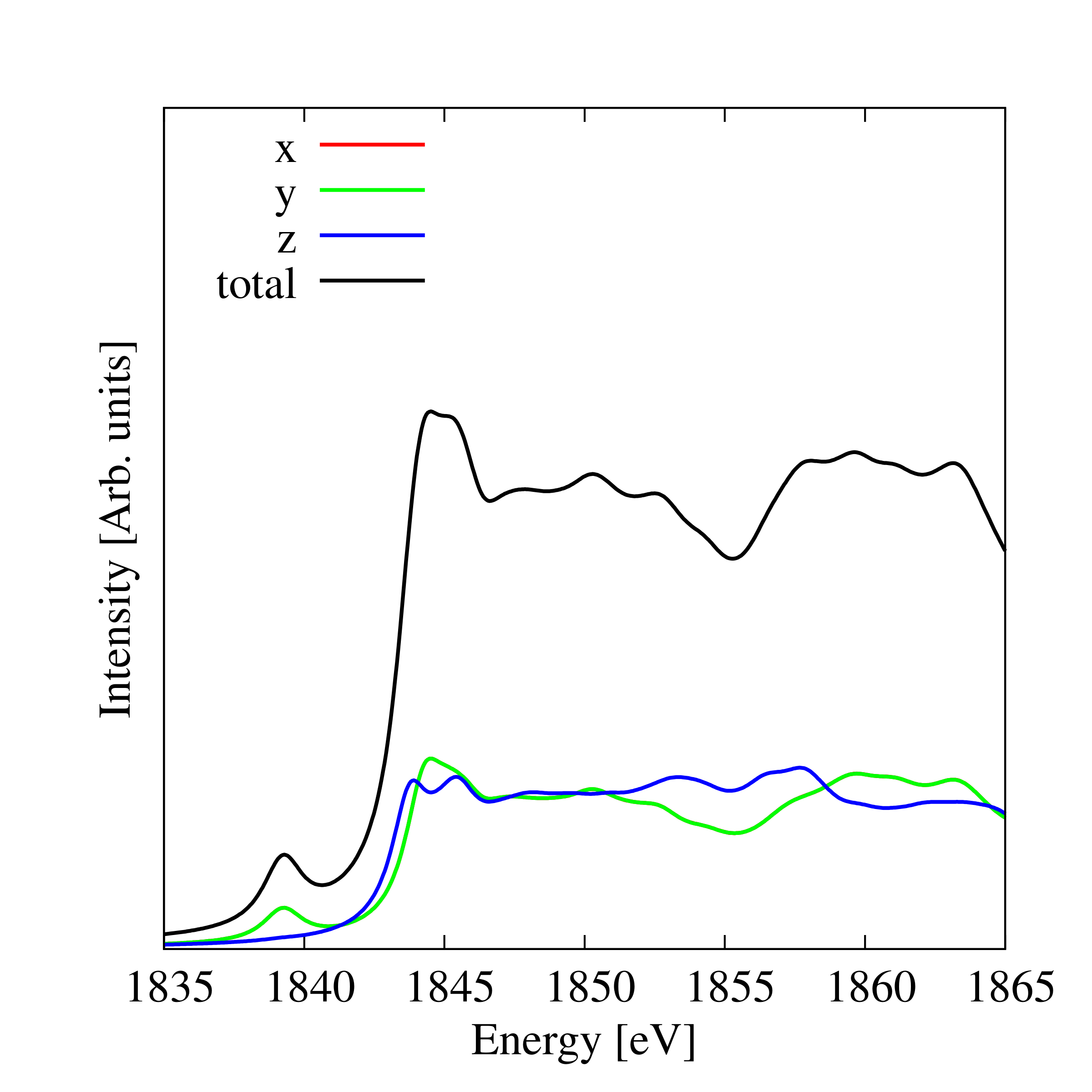}
\caption{Theoretical Si K-edge XANES spectra of a 3$\times$3 supercell of hydrogenated penta-SiC$_2$ with the core-hole at site 1 for $x$-, $y$- and $z$-polarized electric fields. We note that the spectrum for $x$-polarized X-rays (red line, not visible in the diagram) is the same as that for $y$-polarized light (green line).}
\label{fig_PentaHydrogenSiCSiliconTriple}
\end{figure}

The last system for which we have calculated the X-ray absorption is hydrogenated penta-SiC$_2$. Starting with silicon, we report the calculated Si K-edge XANES spectra for hydrogenated penta-SiC$_2$ (3$\times$3 supercell) in Fig. \ref{fig_PentaHydrogenSiCSiliconTriple} with the core-hole in a four-coordinated silicon (site 1 in Fig. \ref{fig_Structures}A).

The similarity with the one case of silicon substitution reported in Fig. \ref{fig_HydrogenatedPentaSiliconTriple} is obvious. In both cases, the local environment of the four-coordinated silicon atom, in which the core-hole is formed, consists of carbon atoms without unsaturated bonds.

\begin{figure}[h!]
\centering
\includegraphics[width=0.5\textwidth]{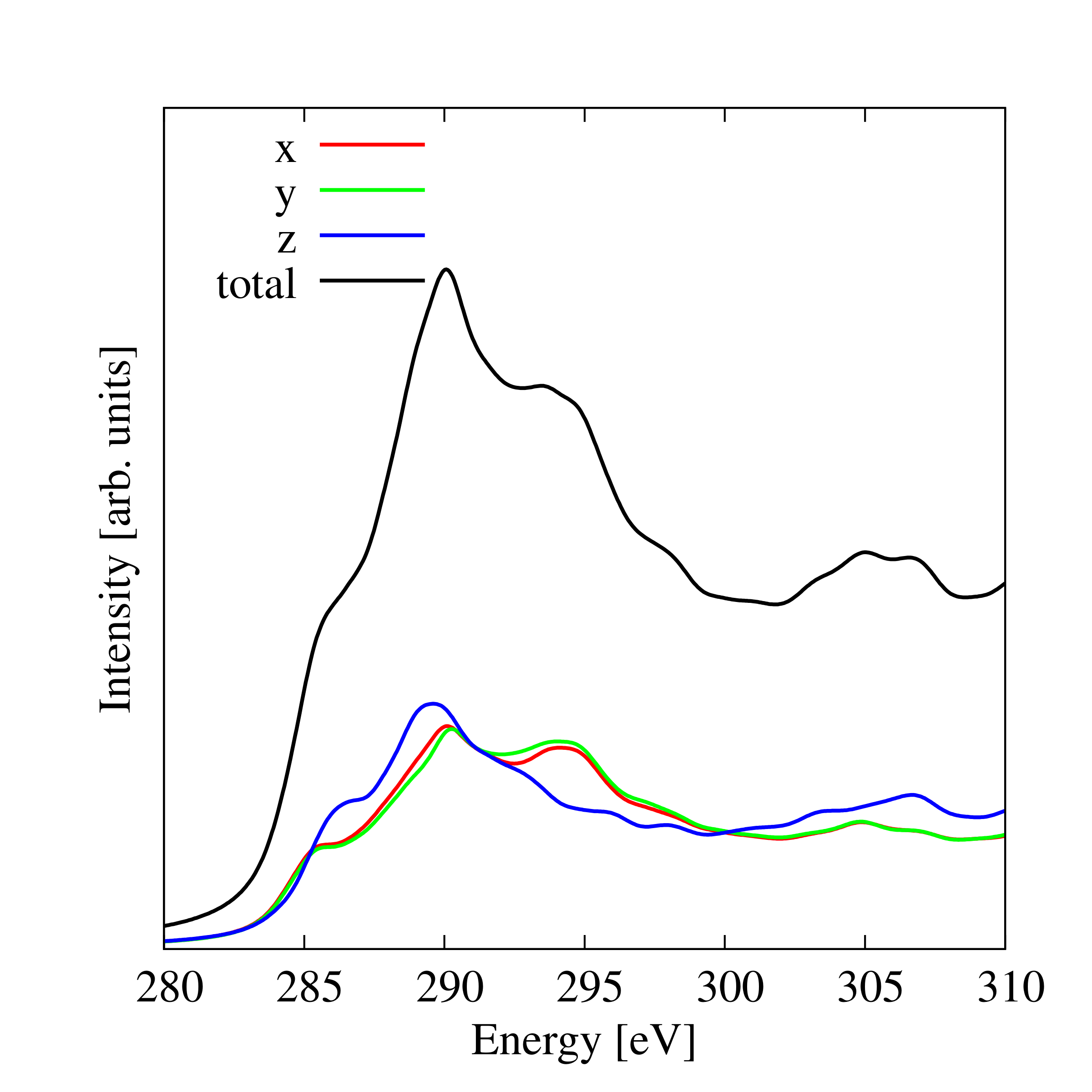}
\caption{Theoretical C K-edge XANES spectra of a 3$\times$3 supercell of hydrogenated penta-SiC$_2$ with the core-hole at site 2 for $x$-, $y$- and $z$-polarized electric fields.}
\label{fig_PentaHydrogenSiCCarbonTriple}
\end{figure}

In Fig. \ref{fig_PentaHydrogenSiCCarbonTriple}, we show the C K-edge spectra for a 3$\times$3 supercell of hydrogenated penta-SiC$_2$ with the core-hole in a three-coordinated position (site 1 in Fig. \ref{fig_Structures}A). In this case, the total spectrum essentially consists of a sharp main peak located at $290$ eV. We also note that the spectra in this case are essentially independent of the polarization direction.

\section{Conclusions}

In this work, we reported a comprehensive study of X-ray absorption for pristine, hydrogenated, and hydroxylated penta-graphene as well as the investigation on the influence of a single silicon substitution in these three cases. In addition, we performed a similar analysis for penta-SiC$_2$, namely for both the pristine and hydrogenated cases. The ab initio XANES spectra calculations were performed within a DFT framework to obtain accurate spectra. The X-ray absorption spectra for different materials show particular features such as peak positions, shapes, and polarization dependencies, which can serve as reliable fingerprints for the experimental identification and discrimination of these structures. Furthermore, these calculations can be used as a basis for the development of the X-ray absorption of these materials.


\vspace{6pt}

\funding{This action has received funding from the European Union under the Mimosa grant agreement No 10104665.
}


\acknowledgments{The authors gratefully acknowledge the use of the HPC facilities at FBK.}


\isPreprints{}{
\begin{adjustwidth}{-\extralength}{0cm}
} 

\reftitle{References}


\bibliography{Bibliografia.bib}


\isPreprints{}{
\end{adjustwidth}
} 
\end{document}